\documentclass[12pt]{article}%
\usepackage{graphics}
\renewcommand{\theequation}{\arabic{section}.\arabic{equation}}%
\begin{document}%
\newcommand{\dd}{{\rm d}}%
\newcommand{\ee}{{\rm e}}%
\newcommand{\ii}{{\rm i}}%
\newcommand{\nab}{\nabla\!}%
\newcommand{\tnab}{\widetilde{\nabla}\!}%
\newcommand{\IR}{{\rm I}\!{\rm R}}%
\def\g{{\mbox{\sl g}}}%
\def\tg{\widetilde{\mbox{\sl g}}}%
\markright{Optical geometry analysis of the electromagnetic
self-force\hfil}%
\title{\bf \LARGE Optical geometry analysis of the electromagnetic
self-force}%
\author{Sebastiano Sonego$^*$ and Marek A.\ Abramowicz$^\dag$$^\ddag$%
\\[4mm]%
{\small\it%
\thanks{\tt sebastiano.sonego@uniud.it}%
\ Universit\`a di Udine, Via delle Scienze 208, 33100 Udine, Italy}%
\\[2mm]%
{\small\it%
\thanks{\tt marek@fy.chalmers.se}%
\ Department of Physics, Chalmers University
of Technology, 41296 G\"oteborg, Sweden}%
\\[2mm]%
{\small\it%
$^\ddag$Nicolaus Copernicus Astronomical Centre,
Polish Academy of Sciences, Warsaw, Poland}%
}%
\date{{\small 16 February 2006;  {\LaTeX-ed \today}}}%
\maketitle%
\begin{abstract}%

We present an analysis of the behaviour of the electromagnetic
self-force for charged particles in a conformally static spacetime,
interpreting the results with the help of optical geometry.  Some
conditions for the vanishing of the local terms in the self-force
are derived and discussed.%

\vspace*{5mm}%
\end{abstract}%

\clearpage



\section{Introduction}%
\label{sec1}%
\setcounter{equation}{0}%

The motion of a point particle with mass $m$, electric charge $e$,
and four-velocity\footnote{We use latin indices $a$, $b$, ... from
the beginning of the alphabet as abstract indices \cite{wald}, while
greek letters $\mu$, $\nu$, ... denote components in some chart and
run from 0 to 3. We work in units in which $G=c=1$, and choose the
positive signature for the metric. We follow the conventions of
Ref.\ \cite{wald} for the curvature tensors; the Ricci tensor
$R_{ab}$ has then the opposite sign than in Refs.\
\cite{dw-b,hobbs}.} $v^ a$, is described by the equation%
\begin{equation}%
m a_a=K_a+F_a\;,%
\label{eqmot}%
\end{equation}%
where $a_a:=v^b\nab_b v_a$ is the particle four-acceleration, $K_a$
is the sum of the external four-forces acting on the particle, and
$F_a$ accounts for the back-reaction of the particle own
electromagnetic field on its motion.  In contrast to a
well-established tradition, we shall avoid referring to $F_a$ as the
{\em radiation reaction\/}, because such a terminology is somehow
improper and can be misleading \cite{rohrlich1,rohrlich2}; we shall
use the term {\em self-force\/} instead.  According to the classic
analysis of DeWitt-Brehme-Hobbs \cite{dw-b,hobbs}, $F_a$ can be
written as the sum of two contributions, $F_a=F^{({\rm
l})}_a+F^{({\rm nl})}_a$, where\footnote{The equation of motion
obtained putting together (\ref{eqmot})--(\ref{nonloc}) contains
third derivatives of the particle coordinates, hence leads to
unphysical conclusions such as preacceleration and runaway
solutions, as it happens for the Lorentz-Dirac equation in flat
spacetime \cite{rohrlich1}. These pathologies can be removed by a
reduction-of-order technique \cite{quinn}.  Nevertheless, in the
following we shall consider the standard expression (\ref{local})
for $F^{({\rm l})}_a$, that differs from the one so obtained only to
higher orders, and is therefore equivalent to it as long as one
limits oneself to the classical domain, in which the theory is
defined.}%
\begin{equation}%
F^{({\rm l})}_a=\frac{2}{3}\,e^2 {k_a}^b v^c\nab_c a_b
+\frac{1}{3}\,e^2 {k_a}^b R_{bc} v^c%
\label{local}%
\end{equation}%
and%
\begin{equation}%
F^{({\rm nl})}_a=e^2v^b\int_{-\infty}^\tau
\dd\tau'\,f_{ab{a'}}v^{a'}\;.%
\label{nonloc}%
\end{equation}%
Here $v^a$ and $\tau$ are the four-velocity and proper time of the
particle, ${k_a}^b={\delta_a}^b+v_a v^b$ is the projector onto the
three-space orthogonal to $v^a$, $R_{ab}$ is the Ricci tensor, and
$f_{ab{a'}}$ is a bi-tensor associated with the presence of
``tails'' in the electromagnetic field \cite{dw-b,tails}. The term
$F^{({\rm l})}_a$ is the relativistic generalisation to a curved
spacetime of the non-covariant expression $(2/3)e^2 \ddot{\bf v}$
for the self-force \cite{rohrlich1,rohrlich2}, and can be regarded
as purely local, as it depends only on quantities evaluated at the
actual position of the particle.  On the contrary, the contribution
$F^{({\rm nl})}_a$ depends on the entire past history of the
particle and on the property of spacetime of being able to
back-scatter electromagnetic waves \cite{dw-b,tails}, encapsulated
in the quantity $f_{ab{a'}}$, and represents therefore an
essentially non-local contribution.  The two terms on the right hand
side of (\ref{local}) are known as the von Laue
\cite{rohrlich1,rohrlich2} and the Hobbs \cite{hobbs} forces,
respectively.  We shall denote them as\footnote{Noteworthy,
the von Laue force can also be written as%
\[%
F^{({\rm vL})}_a=\frac{2}{3}\,e^2\,
\nab_{(v)}^{\:\scriptscriptstyle ({\rm FW})}a_a\;,%
\]%
where $\nab_{(v)}^{\:\scriptscriptstyle ({\rm FW})}$ is the
Fermi-Walker derivative along $v^a$, which for a generic field of
one-forms $u_a$ is \cite{he}%
\[%
\nab_{(v)}^{\:\scriptscriptstyle ({\rm FW})}u_a:=v^b\nab_bu_a+
\left(a_av^b-v_aa^b\right)u_b\;.%
\]%
Hence, the von Laue term $F^{({\rm vL})}_a$ vanishes iff the
acceleration $a_a$ is Fermi-Walker transported along the particle
worldline.}%
\begin{equation}%
F^{({\rm vL})}_a:=\frac{2}{3}\,e^2 {k_a}^b v^c\nab_c a_b%
\label{vonLaue}%
\end{equation}%
and%
\begin{equation}%
F^{({\rm H})}_a:=\frac{1}{3}\,e^2 {k_a}^b R_{bc} v^c\;,%
\label{Hobbs}%
\end{equation}%
so (\ref{local}) can be rewritten as $F^{({\rm l})}_a=F^{({\rm
vL})}_a+F^{({\rm H})}_a$.  It should be clear from the outset that
all this, as well as the treatment that follows, applies to
classical point particles --- an unrealistic model which presents
pathologies, given that the size of a classical object cannot be
smaller than its Compton wavelength. The extension to particles of a
finite size has been discussed
recently \cite{yaghjian}.%

In the last decade there has been a renewed interest in calculations
of self-forces of different nature.  The main motivation for such a
revival is the possibility to go beyond the test particle
approximation when studying motion in a gravitational field, which
is important for investigations about the generation of
gravitational waves.  (For recent reviews, see
\cite{grav-self-force} and references therein.) However, due to the
complexity of the calculations in the gravitational case, several
authors have considered simpler models involving the scalar and
electromagnetic self-interactions, which are also interesting by
themselves.  In particular, much attention has been paid to the
non-local contribution.  In this article we shall instead focus on
the behaviour of the local term $F_a^{({\rm l})}$ for the
electromagnetic case, arguing that it can be understood more easily
in terms of the so-called optical geometry, rather than of the usual
geometry of spacetime.  A similar analysis could be carried out for
scalar and gravitational self-forces, but we shall not cover such
cases here.  Also, we leave for further investigation the issue of
whether optical geometry could be successfully applied to the
challenging problem of computing the non-local term (see, however,
the end of Sec.\ \ref{sec4} for a brief discussion of this point).%

We begin by considering, in the next section, a situation in which
the charge moves uniformly on a circular orbit in a static,
spherically symmetric spacetime, comparing the results in the
special cases of Einstein's universe and Schwarz\-schild spacetime.
For the latter, we shall see that $F_a^{({\rm l})}=0$ at the closed
photon orbit $r=3M$.  This property will be interpreted in Sec.\
\ref{sec3'} by introducing the notion of optical geometry, in which
light paths on $t=\mbox{const}$ hypersurfaces are geodesics.  In
Sec.\ \ref{sec3} we generalise the analysis to an arbitrary
conformally static spacetime, showing that if a charge moves with
constant speed along a possible light path, the self-force is the
one associated with optical geometry (apart from an obvious
conformal rescaling), and providing other conditions for the
different parts of the self-force to vanish.  Section \ref{sec4}
contains a brief summary of the results.%


\section{Uniform circular motion in static spherically
symmetric spacetimes}%
\label{sec2}%
\setcounter{equation}{0}%

Let us begin by discussing a rather special, but highly significant
case. Consider a charge moving on a circular orbit $r=\mbox{const}$
in the plane $\theta=\pi/2$ of a static, spherically symmetric
spacetime, with metric%
\begin{equation}%
\g=-\ee^{2\Phi(r)}\dd t^2+\alpha(r)\dd r^2+
r^2\left(\dd\theta^2+ \sin^2\theta\,\dd\varphi^2\right)\;,%
\label{schw}%
\end{equation}%
where $\Phi$ and $\alpha$ are known functions, with $\alpha$
positive. The components of the four-velocity are%
\begin{equation}%
v^\mu=\Gamma\left(\delta^\mu_t+\Omega\,
\delta^\mu_\varphi\right)\;,%
\label{vmu}%
\end{equation}%
where $\Omega$ is a parameter and
$\Gamma=\left(\ee^{2\Phi}-\Omega^2r^2\right)^{-1/2}$ is fixed by the
normalization $v_\mu v^\mu=-1$. It is convenient to introduce the
velocity $v=\ee^{-\Phi}\Omega r$ measured by static observers, and
the corresponding Lorentz factor
$\gamma:=\left(1-v^2\right)^{-1/2}=\ee^\Phi \Gamma$.%

If the motion is uniform, i.e.\ $\Omega=\mbox{const}$, the
only non-vanishing component of the acceleration is%
\begin{equation}%
a_r=\Gamma^2\left(\ee^{2\Phi}\frac{\dd\Phi}{\dd r}- \Omega^2
r\right)\;,%
\label{agen}%
\end{equation}%
and the von Laue force is given by%
\begin{equation}%
F_\mu^{({\rm vL})}=\frac{2}{3}\,e^2\gamma^5\Omega\, r\,
\frac{\ee^{-\Phi}}{\alpha}\left(1-r\frac{\dd\Phi}{\dd r}\right)
\left(\ee^{2\Phi}\frac{\dd\Phi}{\dd r}-\Omega^2 r\right)
\left(-\Omega\,\delta_\mu^t+\delta_\mu^\varphi\right)\;.%
\label{vLgen}%
\end{equation}%
The Hobbs term is%
\begin{equation}%
F^{({\rm H})}_\mu=\frac{1}{3}\,e^2\gamma^3\Omega\,\ee^{-\Phi}
\left(\ee^{-2\Phi}r^2\,R_{tt}+R_{\varphi\varphi}\right)
\left(-\Omega\,\delta_\mu^t+\delta_\mu^\varphi\right)\;,%
\label{hobbsnew}%
\end{equation}%
where%
\begin{equation}%
R_{tt}=\frac{\ee^{2\Phi}}{\alpha}\left(\frac{\dd^2\Phi}{\dd r^2}
+\frac{2}{r}\frac{\dd\Phi}{\dd r}+\left(\frac{\dd\Phi}{\dd
r}\right)^2-\frac{1}{2\,\alpha}\frac{\dd\Phi}{\dd
r}\frac{\dd\alpha}{\dd r}\right)%
\label{Rtt}%
\end{equation}%
and%
\begin{equation}%
R_{\varphi\varphi}=1-\frac{1}{\alpha}
+\frac{r}{\alpha}\left(\frac{1}{2\,\alpha}\frac{\dd\alpha}{\dd
r}-\frac{\dd\Phi}{\dd r}\right)%
\label{Rphiphi}%
\end{equation}%
are the only relevant non-vanishing components of the Ricci
tensor.\footnote{The component $R_{\varphi\varphi}$ in
(\ref{Rphiphi}) is evaluated at $\theta=\pi/2$.  The general
expression for $R_{\varphi\varphi}$ contains an overall extra
coefficient $\sin^2\theta$.}  Because of the rather complicated
dependence on $\Phi$, $\alpha$, and their derivatives, it is more
instructive to focus on two particular cases.%


\subsection{Einstein's universe}%
\label{sec2a}%

\begin{figure}[htbp]
\vbox{ \hfil \scalebox{0.50}{{\includegraphics{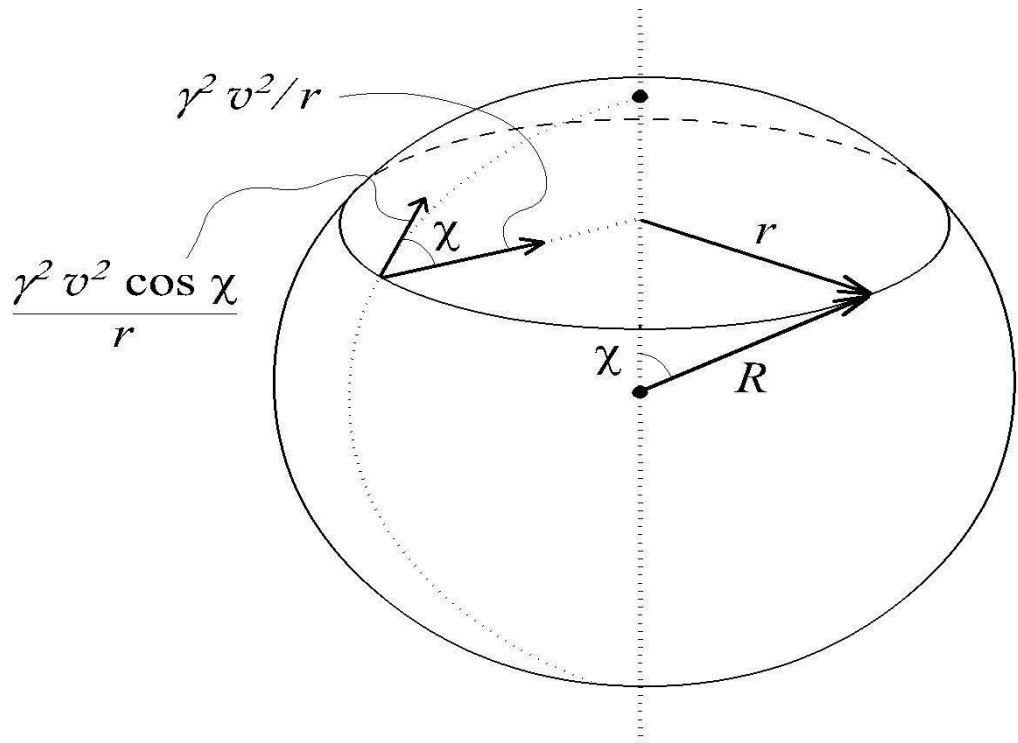}}} \hfil }
\vskip1cm%
\caption{Embedding diagram for the surface $t=\mbox{const}$,
$\theta=\pi/2$ in Einstein's universe, showing the geometrical
meaning of the quantities $R$, $r$, and $\chi$.  The centripetal
acceleration of a particle moving at $r=\mbox{const}$ is given by
the tangential component of the acceleration in the fictitious
three-dimensional Euclidean space.  For $\chi<\pi/2$ the centripetal
acceleration points in the direction of decreasing $r$, while for
$\pi/2<\chi<\pi$ it points in the direction of increasing $r$.}%
\label{fig0}%
\end{figure}

The metric of Einstein's static universe corresponds to $\Phi=0$ and
$\alpha(r)=\left(1-r^2/R^2\right)^{-1}$, where $R$ is a positive
parameter (the ``radius'' of the three-dimensional spherical space).
The coordinate $r\in[\,0,R\,[\,$, however, does not cover the whole
manifold, so it is convenient to introduce a new variable
$\chi\in[\,0,\pi\,]$ defined through $r=R\sin\chi$.  The metric then
takes the form%
\begin{equation}%
\g=-\dd t^2+R^2\dd\chi^2+R^2\sin^2\chi\left(\dd\theta^2+
\sin^2\theta\,\dd\varphi^2\right)\;,%
\end{equation}%
showing that $R\chi$ measures the proper distance from the point
$\chi=0$. The function $\alpha$ becomes
$\alpha(\chi)=1/\cos^2\chi$, and the only non-vanishing component
of the acceleration is%
\begin{equation}%
a_\chi=\frac{\partial r}{\partial\chi}a_r=
-\frac{\Omega^2R^2\sin\chi\cos\chi}{1-\Omega^2R^2\sin^2\chi}\;.%
\end{equation}%
The magnitude (with sign) of the acceleration is
$\gamma^2(v^2/r)\cos\chi$, so the acceleration differs from the
special relativistic expression $\gamma^2v^2/r$ only by the factor
$\cos\chi$, whose origin can be easily understood by looking at
Fig.\ \ref{fig0}.  %
Note that the acceleration for orbits with
$\chi\in\,]\,\pi/2,\pi\,[\,$ has the opposite sign than for those
with $\chi\in\,]\,0,\pi/2\,[\,$, and vanishes when $\chi=\pi/2$.%

The von Laue force is%
\begin{equation}%
F_\mu^{({\rm vL})}=\frac{2}{3}\,e^2\gamma^5\Omega^3R^2
\sin^2\chi\cos^2\chi\left(-\Omega\,\delta_\mu^t+
\delta_\mu^\varphi\right)\;.%
\end{equation}%
It also vanishes for $\chi=\pi/2$, but does not exhibit any
change in sign.%

Finally, let us compute the Hobbs force. The components of the Ricci
tensor can be evaluated directly from the expressions (\ref{Rtt})
and (\ref{Rphiphi}), or simply remembering that the symmetries of
Einstein's universe imply $R_{\mu\nu}=(2/R^2)h_{\mu\nu}$, where%
\begin{eqnarray}%
h&=&\left(1-r^2/R^2\right)^{-1}\dd r^2+
r^2\left(\dd\theta^2+\sin^2\theta\,\dd\varphi^2\right)\nonumber\\%
&=&R^2\dd\chi^2+R^2\sin^2\chi\left(\dd\theta^2+
\sin^2\theta\,\dd\varphi^2\right)%
\end{eqnarray}%
is the metric tensor on the $t=\mbox{const}$ hypersurfaces
\cite{weinberg}. Thus, we have%
\begin{equation}%
F_\mu^{({\rm H})}=\frac{2}{3}\,e^2\gamma^3\Omega
\sin^2\chi\left(-\Omega\,\delta_\mu^t+
\delta_\mu^\varphi\right)\;.%
\end{equation}%
This expression is non-vanishing for all non-trivial values of
$\chi$. Therefore, the local part of the self-force,%
\begin{equation}%
F_\mu^{({\rm l})}=\frac{2}{3}\,e^2\gamma^5\Omega
\sin^2\chi\left(1+\Omega^2 R^2\cos
2\chi\right)\left(-\Omega\,\delta_\mu^t+
\delta_\mu^\varphi\right)\;,%
\end{equation}%
never vanishes, in spite of the fact that when the motion takes
place on the spatial geodesic $\chi=\pi/2$, the von Laue force
does.%


\subsection{Schwarzschild spacetime}%
\label{sec2b}%

In Schwarz\-schild spacetime,%
\begin{equation}%
\Phi=\frac{1}{2}\,\ln\left(1-\frac{2M}{r}\right)%
\label{Phisch}%
\end{equation}%
and $\alpha=\ee^{-2\Phi}$. Then, from Eq.\ (\ref{agen}) we
obtain%
\begin{equation}%
a_r=\Gamma^2\left(\frac{M}{r^2}-\Omega^2\,r\right)=
\frac{M-\Omega^2\,r^3}{r(r-2M-\Omega^2\,r^3)}%
\label{amu}%
\end{equation}%
and, from (\ref{vLgen}),%
\begin{equation}%
F_\mu^{({\rm vL})}=\frac{2}{3}\,e^2\gamma^5\Omega
r\left(1-\frac{2M}{r}\right)^{-3/2}\left(1-
\frac{3M}{r}\right)\left(\frac{M}{r^2}-\Omega^2 \,r\right)
\left(-\Omega\,\delta_\mu^t+\delta_\mu^\varphi\right)\;.%
\label{dmu}%
\end{equation}%
Since $R_{ab}=0$, the Hobbs force vanishes identically, so
$F_\mu^{({\rm l})}=F_\mu^{({\rm vL})}$.%

Let us now study the behaviour of $a_r$ and $F_\mu^{({\rm
l})}$ for different values of $r$. When $r/M$ is very large
(Newtonian gravity), Eq.\ (\ref{amu}) can be approximated as%
\begin{equation}%
a_r\approx \gamma^2 \left(\frac{M}{r^2}-
\Omega^2 r\right)\;.%
\label{amu1}%
\end{equation}%
Apart from the factor $\gamma^2$, which represents only
special-relativistic corrections, the norm of $a_a$ is just the
difference between the gravitational and the centripetal
accelerations in Newtonian mechanics. This agrees with the physical
interpretation of $a_a$, which represents the thrust per unit mass
that must be applied to the particle in order to keep it at the
fixed value of $r$ with angular velocity $\Omega$. In particular,
$a_a=0$ for a Keplerian motion, $\Omega^2=\Omega_{\scriptscriptstyle
K}(r)^2:=M/r^3$ or $r=r_{\scriptscriptstyle
K}(\Omega):=(M/\Omega^2)^{1/3}$. In the same approximation,%
\begin{equation}%
F_\mu^{({\rm l})}\approx\frac{2}{3}\,e^2\Omega
r\gamma^5\left(\frac{M}{r^2}-\Omega^2
r\right)\left(-\Omega\,\delta_\mu^t+
\delta_\mu^\varphi\right)\;.%
\label{dmu1}%
\end{equation}%
The local part of the self-force behaves then in the following
way: When the motion is Keplerian it vanishes, consistently with
the fact that $a_a=0$; in the super-Keplerian regime ($\Omega^2>
\Omega_{\scriptscriptstyle K}(r)^2$ or $r>r_{\scriptscriptstyle
K}(\Omega)$) it points backward with respect to the direction of
motion; finally, in the sub-Keplerian regime
($\Omega^2<\Omega_{\scriptscriptstyle K}(r)^2$ or
$r<r_{\scriptscriptstyle K}(\Omega)$) it points forward with
respect to the direction of motion. In the particular case of a
static charge, $\Omega=0$ and $F_\mu^{({\rm l})}=0$.%

It is interesting to note that this behaviour is peculiar to gravity
and has no counterpart for motion in other central fields such as,
e.g., a Coulomb one. This is easily seen by computing the
self-force associated with circular motion in Minkowski spacetime,%
\begin{equation}%
F_\mu\equiv F_\mu^{({\rm l})}=-\frac{2}{3}\,e^2\Omega^3
r^2\gamma^5\left(-\Omega\,\delta_\mu^t+
\delta_\mu^\varphi\right)\;,%
\label{coul}%
\end{equation}%
where the second equality follows immediately from (\ref{dmu}) with
$M=0$ (see also Ref.\ \cite{burko}, where this expression is derived
explicitly for the case of synchrotron radiation). Now, the
self-force always points backward with respect to the direction of
motion, regardless of the magnitude of $\Omega$. This is related to
the fact that the right hand side of Eq.\ (\ref{coul}) does not
contain any indication about the central force responsible for
keeping the charge on circular motion, contrary to what happens in
the gravitational case, where a term $M/r^2$ appears together with
$\Omega^2 r$.  The combination $M/r^2-\Omega^2r$ is the same that
appears in $a_r$ --- see Eq.\ (\ref{amu}) ---, and is ultimately
linked to the association between gravity and curvature. Thus,
although for most purposes one can think of Newtonian gravity as a
field on a flat background, the present analysis shows that this
picture would lead to incorrect conclusions as far as the self-force
is concerned.\footnote{Apparently, this point is not appreciated in
the extant literature. See, e.g., Ref.\ \cite{rohrlich1}, p.\ 183,
for an explicit statement that the Newtonian and Coulomb
problems are alike.}%

Let us come back to the analysis of $a_r$ and $F_\mu^{({\rm l})}$.
For $r>3M$ the qualitative behaviour of these quantities at
different values of $r$ is like in the Newtonian limit examined
above, even when $r/M$ is not large. However, it is easily seen from
Eqs.\ (\ref{amu}) and (\ref{dmu}) that when $r=3M$, $a_r=(3M)^{-1}$
and $F_\mu^{({\rm l})}=0$. Thus, the thrust needed in order to keep
a particle on the closed photon orbit does not depend on the
particle speed \cite{al}. Furthermore, if the particle is charged,
the local part of the self-force vanishes, in spite of the fact that
motion takes place on a circle and is not geodesic.\footnote{This
point is related to earlier results \cite{sokolov}.} Finally, for
orbits with $2M<r<3M$, it follows from (\ref{dmu}) that the local
part of the self-force always points backward with respect to the
direction of motion. Since such orbits are necessarily
sub-Keplerian, as it follows from the inequality\footnote{Proof:
Consider the condition $2M/r+\Omega^2r^2<1$, which guarantees that
$\Gamma$ be real or, equivalently, that the particle worldline be
timelike; subtract $\Omega_{\scriptscriptstyle K}(r)^2 r^2$ from
both sides, and use the definition
$\Omega_{\scriptscriptstyle K}(r)^2:=M/r^3$.}%
\begin{equation}%
\left(\Omega^2-\Omega_{\scriptscriptstyle
K}(r)^2\right)r^2<1-3M/r\;,%
\end{equation}%
this behaviour is the opposite than for $r>3M$.%


\section{Optical geometry}%
\label{sec3'}%

The behaviour of $a_r$ can be understood in the following way. The
component $a_r$ of the acceleration, given by the expression
(\ref{amu}), is proportional to the thrust necessary to keep the
particle on its orbit, and can be written as the difference between
a gravitational part $a_r^{({\rm g})}$, independent of $\Omega$, and
a centripetal part $a_r^{({\rm c})}$:%
\begin{equation}%
a_r^{({\rm g})}=\frac{M}{r^2\left(1-2M/r\right)}\;;%
\end{equation}%
\begin{equation}%
a_r^{({\rm c})}=\Omega^2 r\,\frac{r-3M}{\left(1-
2M/r\right)\left(r-2M-\Omega^2r^3\right)}\;.%
\label{ag}%
\end{equation}%
(Of course, $a_r^{({\rm g})}$ and $a_r^{({\rm c})}$ have signs
opposite to those of the gravitational and centripetal forces.
Alternatively, one can identify $a_r^{({\rm c})}$ with the
centrifugal field acting on the particle in the comoving frame.) It
is then evident from (\ref{ag}) that the centripetal force vanishes
at $r=3M$, for which $a_r=a_r^{({\rm g})}=(3M)^{-1}$. In addition,
(\ref{ag}) predicts that $a_r^{({\rm c})}$ has opposite sign in the
regions $r>3M$ and $r<3M$.%

\begin{figure}[htbp]
\vbox{ \hfil \scalebox{0.50}{{\includegraphics{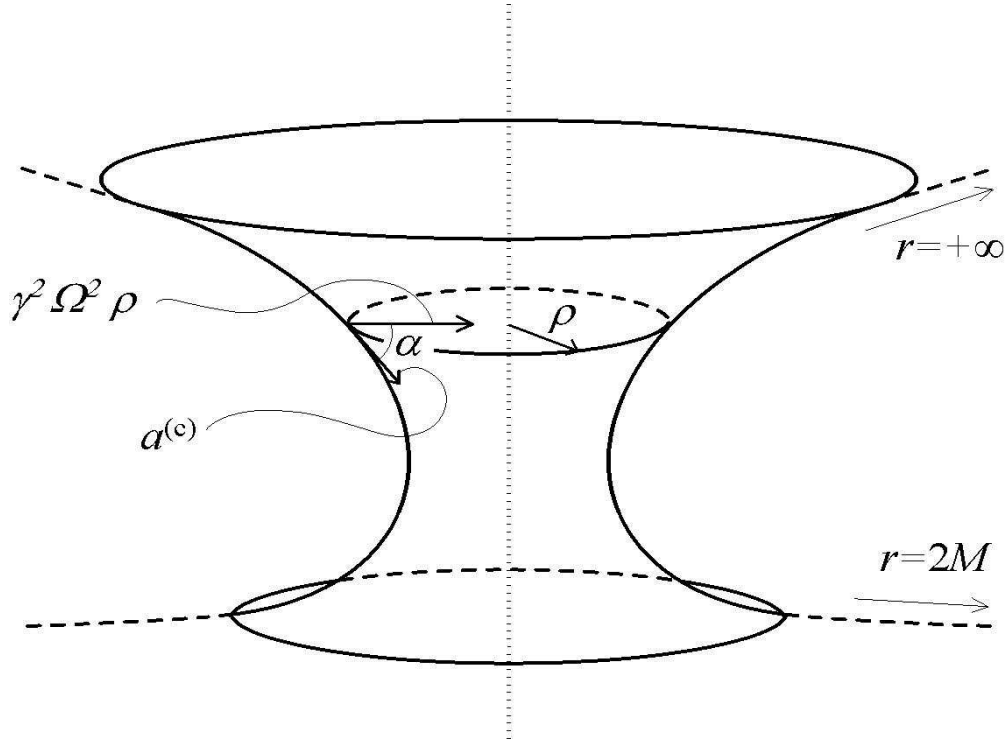}}} \hfil }
\vskip1cm%
\caption{Embedding diagram for part of the surface $t=\mbox{const}$,
$\theta=\pi/2$ in Schwarz\-schild spacetime, with the optical
geometry. The throat corresponds to $r=3M$. The centripetal
acceleration of a particle moving at $r=\mbox{const}$ is given by
the tangential component of the acceleration in the fictitious
three-dimensional Euclidean space. Notice that for $r>3M$ the
centripetal acceleration points in the direction of decreasing $r$,
while for $2M<r<3M$ it points in the direction of increasing $r$.}%
\label{fig1}%
\end{figure}

These properties admit a simple explanation if one imagines that the
particle motion takes place in the so-called optical spacetime
\cite{acl}, with a metric $\tg=\left(1-2M/r\right)^{-1}\g$, under
the action of a ``gravitational potential'' $\Phi$ given by
(\ref{Phisch}), which produces a ``gravitational field''
\cite{a93}%
\begin{equation}%
\tilde{g}_\mu=-{k_\mu}{}^\nu\nab_\nu\Phi=
-\left(1-\frac{2M}{r}\right)^{-1}\frac{M}{r^2}\,\delta_\mu^r\;.%
\end{equation}%
We have then $a_r^{({\rm g})}=-\tilde{g}_r$ and, in the optical
spacetime, the magnitude of the gravitational field is given by
the simple Newtonian expression $\left(\tg^{\mu\nu}\tilde{g}_\mu
\tilde{g}_\nu\right)^{1/2}=M/r^2$. Also, defining
$\tilde{v}^\mu:=(1-2M/r)^{1/2} v^\mu$, so that
$\tg_{\mu\nu}\tilde{v}^\mu\tilde{v}^\nu=-1$, one gets%
\begin{equation}%
\tilde{a}_\mu=\tg_{\mu\sigma}
\tilde{v}^\nu\tnab_\nu\tilde{v}^\sigma=
-\frac{\Gamma^2\Omega^2(r-3M)}{1-2M/r}\delta_\mu^r\;,%
\end{equation}%
and $a_r^{({\rm c})}=\tilde{a}_r$. The magnitude of the
acceleration in optical spacetime is%
\begin{equation}%
\left(\tg^{\mu\nu}\tilde{a}_\mu
\tilde{a}_\nu\right)^{1/2}=\Gamma^2\Omega^2|r-3M|\;.%
\label{optacc}%
\end{equation}%
The presence of the factor $(r-3M)$ in $a_r^{({\rm c})}$ can be
understood intuitively \cite{a90} considering an embedding diagram
\cite{ksa} of the section $\theta=\pi/2$ of the optical space
$({\cal S},\tilde{h}_{ab})$, where $\cal S$ is any $t=\mbox{const}$
hypersurface of the Schwarz\-schild spacetime, and the metric
$\tilde{h}_{ab}$ has the coordinate representation \cite{acl}%
\begin{equation}%
\tilde{h}=\left(1-\frac{2M}{r}\right)^{-2}\dd r^2+
\left(1-\frac{2M}{r}\right)^{-1}
r^2\left(\dd\theta^2+\sin^2\theta\,\dd\varphi^2\right)%
\label{orgschw}%
\end{equation}%
(see Fig.\ \ref{fig1} and the Appendix).%

Let us now consider the self-force. The von Laue term associated
with the optical quantities has the components%
\begin{equation}%
\widetilde{F}_\mu^{({\rm
vL})}=-\frac{2}{3}\,e^2\gamma^5\Omega^3r^2\left(1-\frac{2M}{r}\right)^{-2}
\left(1-\frac{3M}{r}\right)^2\left(-\Omega\,\delta_\mu^t+
\delta_\mu^\varphi\right)\;,%
\label{vonL}%
\end{equation}%
so it always points backward with respect to the direction of
motion, vanishes only at $r=3M$, and reduces to (\ref{coul}) in the
region $r\gg M$. The Hobbs force in the optical space can be
computed either directly, or by using the transformation formula
(2.11) of Ref.\ \cite{s98}. Its components are%
\begin{equation}%
\widetilde{F}_\mu^{({\rm
H})}=\frac{2}{3}\,e^2\gamma^3\Omega\,\frac{M}{r}\left(1-
\frac{2M}{r}\right)^{-1}
\left(1-\frac{3M}{r}\right)\left(-\Omega\,\delta_\mu^t+
\delta_\mu^\varphi\right)\;.%
\label{hobbs}%
\end{equation}%
For $r\gg M$, this becomes%
\begin{equation}%
\widetilde{F}_\mu^{({\rm
H})}\approx\frac{2}{3}\,e^2\gamma^3\Omega\,\frac{M}{r}
\left(-\Omega\,\delta_\mu^t+\delta_\mu^\varphi\right)\;.%
\end{equation}%
Note that the magnitude of the von Laue force is proportional to
$\Omega^3 r$, like in Minkowski space, while the magnitude of the
Hobbs force is proportional to $M\Omega/r^2$, which can be rewritten
as $M/r^3$ times the speed $\Omega r$. Thus, from the
perspective offered by optical geometry, the coefficient%
\[ \frac{M}{r^2}-\Omega^2 r \]%
that appears in Eq.\ (\ref{dmu1}) should not be regarded as
representing the difference between the gravitational and the
centripetal forces, because the first term actually arises from the
{\em curvature\/} of $({\cal S},\tilde{h}_{ab})$.  The same
coefficient is present for uniform circular motion at
$r=\mbox{const}$ on $({\cal S},\tilde{h}_{ab})$, regardless of the
nature (gravitational or other) of the ``central'' force. Therefore,
when the local part of the self-force is analysed in optical
geometry, the puzzling peculiarity of gravity that was
noticed in Sec.\ \ref{sec2b} disappears.%

It is worth pointing out that $F_\mu^{({\rm l})}$,
$\widetilde{F}_\mu^{({\rm vL})}$, and $\widetilde{F}_\mu^{({\rm
H})}$ all vanish at $r=3M$, as it follows from Eqs.\ (\ref{dmu}),
(\ref{vonL}) and (\ref{hobbs}). However, while the property
$\widetilde{F}_\mu^{({\rm vL})}(3M)=0$ is natural from the point of
view of optical geometry, because the circle $r=3M$ is a geodesic of
$({\cal S},\tilde{h}_{ab})$ and the charge motion is uniform, the
fact that also $\widetilde{F}_\mu^{({\rm H})}(3M)=0$
--- hence $F_\mu^{({\rm l})}(3M)=0$ --- is due to the algebraic
properties of the optical Ricci tensor at $r=3M$ rather than to the
geodesic character in $({\cal S},\tilde{h}_{ab})$ of the closed
photon orbit (see Sec.\ \ref{sec3b} below). Therefore, the change of
sign of $F_\mu^{({\rm l})}$ at $r=3M$ has a different origin than
for other phenomena in which a similar ``reversal'' takes place
\cite{rev}.%


\section{General results}%
\label{sec3}%
\setcounter{equation}{0}%

If $({\cal M},\g_{ab})$ and $({\cal M}, \tg_{ab})$ are two
conformally related spacetimes, with%
\begin{equation}%
\tg_{ab}=\ee^{-2\Phi}\g_{ab}\;,%
\label{transf}%
\end{equation}%
the local and non-local parts of the corresponding self forces are
related as $F^{({\rm l})}_a=\ee^{-\Phi} \widetilde{F}^{({\rm l})}_a$
and $F^{({\rm nl})}_a=\ee^{-\Phi} \widetilde{F}^{({\rm nl})}_a$
\cite{s98}. Then, in order to compute $F^{({\rm l})}_a$ or $F^{({\rm
nl})}_a$, one can work in a conformally related spacetime $({\cal
M}, \tg_{ab})$ rather than in the physical one, $({\cal M},
\g_{ab})$.  We now use this remark to put the results of Sec.\
\ref{sec2} into a more general context extending the line of
reasoning outlined in Sec.\ \ref{sec3'}, and to establish some extra
necessary and/or sufficient conditions for the different parts
of $F^{({\rm l})}_a$ to vanish.%

Let us consider a charge with four-velocity $v^a$ in an arbitrary
conformally static\footnote{This restriction is motivated mainly by
simplicity.  Optical geometry can be straightforwardly defined also
for conformally stationary spacetimes \cite{torres}, and has been
extended even to cases where no global time symmetry exists, as it
happens in gravitational collapse \cite{saa}.  For recent results
concerning generalisations of optical geometry, see Ref.\
\cite{rickard}.} spacetime $({\cal M},\g_{ab})$. By definition, such
a spacetime admits a hypersurface-orthogonal timelike conformal
Killing vector field $\eta=\partial/\partial t$, corresponding to a
conformal Killing time $t$.  One can then define the scalar
function%
\begin{equation}%
\Phi:=\frac{1}{2}\,\ln\left(-\g_{ab}\eta^a\eta^b\right)%
\label{Phi}%
\end{equation}%
and consider the ultrastatic spacetime $({\cal M}, \tg_{ab})$ with
$\tg_{ab}$ given by (\ref{transf}), in which $\eta^a$ is a Killing
vector field with unit norm, $\pounds_\eta \tg_{ab}=0$ and
$\tg_{ab}\eta^a\eta^b=-1$ \cite{fulling}.  It is not difficult to
check that the projection, along the integral curves of $\eta^a$, of
null geodesics of $({\cal M},\g_{ab})$ onto the spatial
hypersurfaces $\cal S$ defined by the condition $t=\mbox{const}$,
are geodesics of $({\cal S},\tilde{h}_{ab})$, where
$\tilde{h}_{ab}:=\tg_{ab}+\eta_a\eta_b$ and
$\eta_a:=\tg_{ab}\eta^b$. Hence, generalising what we have done in
Sec.\ \ref{sec3'}, all quantities pertaining to the spacetime
$({\cal M},\tg_{ab})$ will be denoted as ``optical.'' Instead of
$F^{({\rm l})}_a$, we can then consider $\widetilde{F}^{({\rm
l})}_a$, computed in the optical spacetime $({\cal M},\tg_{ab})$
starting from the rescaled four-velocity $\tilde{v}^a=\ee^\Phi v^a$.%

At any point on the particle worldline, $\tilde{v}^a$ can be
decomposed in a unique way as%
\begin{equation}%
\tilde{v}^a=\gamma\left(\eta^a+v\tilde{\tau}^a\right)\;,%
\label{dec}%
\end{equation}%
where $v\in\,]-1,1[\,$ is the speed of the particle according to an
observer with four-velocity parallel to $\eta^a$, the coefficient
$\gamma:=\left(1- v^2\right)^{-1/2}$ is the corresponding Lorentz
factor, and $\tilde{\tau}^a$ is a spacelike vector orthogonal to
$\eta^a$, such that $\tg_{ab}\tilde{\tau}^a \tilde{\tau}^b=1$.  The
four-acceleration $\tilde{a}_a:=\tilde{v}^b\tnab_b \tilde{v}_a$ of
the particle in the optical spacetime $({\cal M},\tg_{ab})$ is then%
\begin{equation}%
\tilde{a}_a=\gamma^2\widetilde{T}_a\tilde{v}^b\tnab_b v
+\gamma^2v^2\tilde{\tau}^b\tnab_b\tilde{\tau}_a+ \gamma^2
v\eta^b\tnab_b\tilde{\tau}_a\;,%
\label{atilde}%
\end{equation}%
where%
\begin{equation}%
\widetilde{T}^a:=\frac{{\tilde{k}^a}{}_b\tilde{\tau}^b}
{\sqrt{\tilde{k}_{cd}\tilde{\tau}^c\tilde{\tau}^d}}=
\gamma\left(\tilde{\tau}^a+v\,\eta^a\right)%
\end{equation}%
is a unit vector in $({\cal M},\tg_{ab})$, orthogonal to
$\tilde{v}^a$.  Using now the orthogonality condition
$\eta^a\tilde{\tau}_a =0$, and the property $\tnab_b\eta_a=0$ that
follows from the fact that $({\cal M},\tg_{ab})$ is ultrastatic,
we have%
\begin{equation}%
\tnab_b\tilde{\tau}_a=\widetilde{\rm D}_b\tilde{\tau}_a
-\eta_b\eta^c\tnab_c\tilde{\tau}_a\;,%
\label{nablatau}%
\end{equation}%
where $\widetilde{\rm D}_b$ is the covariant derivative along
directions orthogonal to $\eta^a$, hence on $({\cal
S},\tilde{h}_{ab})$.  In order for both Eqs.\ (\ref{atilde}) and
(\ref{nablatau}) to make sense, one needs to know $\tilde{\tau}^a$
off the particle worldline. The natural extension of
$\tilde{\tau}^a$ is such that its integral curves (located on
$t=\mbox{const}$ hypersurfaces) coincide with the projection on
$\cal S$ of the particle worldline along the integral curves of
$\eta^a$.  This is equivalent to requiring that
$\pounds_\eta\tilde{\tau}^a=0$, i.e.\ that
$\eta^b\tnab_b\tilde{\tau}^a=\tilde{\tau}^b\tnab_b\eta^a$, and the
latter quantity vanishes in an ultrastatic spacetime.  Hence,
$\eta^b\tnab_b\tilde{\tau}_a=0$, and the particle acceleration in
optical spacetime takes the very simple form%
\begin{equation}%
\tilde{a}_a=\gamma^2\widetilde{T}_a\tilde{v}^b\tnab_b v+
\gamma^2v^2\tilde{\tau}^b\widetilde{\rm D}_b\tilde{\tau}_a\;,%
\label{a4}%
\end{equation}%
in which one can identify a tangential and an orthogonal
(centripetal) component \cite{a93}.%

The relationship \cite{s98}%
\begin{equation}%
\tilde{a}_a=a_a-{k_a}^b\nab_b\Phi%
\label{relationship}%
\end{equation}%
generalises the result, already obtained in Sec.\ \ref{sec3'} for
the special case of Schwarz\-schild spacetime, according to which
one can write the particle acceleration in optical spacetime,
$\tilde{a}_a$, as the sum between the non-gravitational thrust per
unit mass, $a_a$, and a gravitational force per unit mass,
$\tilde{g}_a=-{k_a}^b\nab_b\Phi$:%
\begin{equation}%
\tilde{a}_a=a_a+\tilde{g}_a\;.%
\label{a=a+g}%
\end{equation}%
Equation (\ref{a=a+g}) contains a precise formulation of the
equivalence principle, because the quantity that is physically
measurable at the particle location --- the thrust, proportional to
$a_a$ --- allows one only to determine the difference
$\tilde{a}_a-\tilde{g}_a$ between $\tilde{a}_a$ and $\tilde{g}_a$,
but not $\tilde{a}_a$ and $\tilde{g}_a$ separately (unless one
possesses extra knowledge, of a non-local nature, about the
structure of the spacetime under consideration).\footnote{In
particular, the two situations with $\tilde{a}_a=a_a$ and
$\tilde{g}_a=0$, or with $\tilde{a}_a=0$ and $\tilde{g}_a=-a_a$,
cannot be distinguished by a measurement of the thrust.}  Inserting
(\ref{a4}) into (\ref{a=a+g}), and identifying
$-\gamma^2\widetilde{T}_a\tilde{v}^b\tnab_b v$ and
$-\gamma^2v^2\tilde{\tau}^b\widetilde{\rm D}_b\tilde{\tau}_a$ with
suitable inertial forces per unit mass \cite{a93,perlick}, one finds
a general relativistic version of the statement, commonly expressed
within the framework of Newtonian mechanics, that in a frame
comoving with the particle there is perfect balance between the
non-gravitational thrust, the gravitational force, and the inertial
forces acting on the particle.%

On replacing (\ref{a4}) into the definition of
$\widetilde{F}_a^{({\rm l})}$ one can find the local part of the
self-force expressed in terms of $v$ and $\tilde{\tau}^a$, that
characterize the motion in optical spacetime $({\cal M},\tg_{ab})$.
However, since the general expression is not particularly
illuminating, let us focus instead on the most interesting
particular cases.%


\subsection{Uniform motion along optical geodesics}%
\label{sec3a}%

It is obvious from Eq.\ (\ref{a4}) that a particle which moves
uniformly (i.e., with $\tilde{v}^b\tnab_b v=0$) along optical
geodesics of space (so that $\tilde{\tau}^b\widetilde{\rm
D}_b\tilde{\tau}_a=0$) has vanishing acceleration in $({\cal
M},\tg_{ab})$, that is, $\tilde{a}_a=0$. Then
$\widetilde{F}_a^{({\rm l})}$ consists only of the Hobbs
term and we can write%
\begin{equation}%
F_a=\ee^{-\Phi}\left(\frac{1}{3}\,e^2{k_a}{}^b
\widetilde{R}_{bc}\tilde{v}^c+e^2\tilde{v}^b
\int_{-\infty}^{\tilde{\tau}}\dd\tilde{\tau}'\,
\tilde{f}_{ab{a'}}\tilde{v}^{a'}\right)\;.%
\label{main}%
\end{equation}%
Thus, the particle is indeed subjected to a self-force, but just
to the one that is associated with the geometric properties of the
optical space. In a sense, the charge ``feels'' the geometry of
$({\cal M},\tg_{ab})$ rather than the one of the physical
spacetime $({\cal M},\g_{ab})$.%

It may be interesting to note that, with the only exception of
ultrastatic spacetimes, for which the ordinary and the optical
geometries simply coincide, there are no lines in the ordinary space
$({\cal S},h_{ab})$ such that the von Laue term vanishes identically
for a charge that moves uniformly along them.  On the contrary, in
$({\cal S},\tilde{h}_{ab})$ the conditions that $v=\mbox{const}$ and
$\widetilde{F}_a^{({\rm vL})}=0$ uniquely select optical geodesics.%


\subsection{Special metrics}%
\label{sec3b}%

If ${k_a}^b\widetilde{R}_{bc}\tilde{v}^c=0$, then the expression
(\ref{main}) implies that $F_a=F_a^{({\rm nl})}$, because
$F_a^{({\rm l})}$ and $F_a^{({\rm nl})}$ do not mix under a
conformal transformation \cite{s98}.\footnote{For a more general,
non-uniform motion, one finds $\widetilde{F}_a^{({\rm
l})}=\widetilde{F}_a^{({\rm vL})}$.} This can happen iff
${\widetilde{R}_a}{}^b\tilde{v}_b=\lambda\tilde{v}_a$ for some
$\lambda$, and it is easy to see, by contracting this equation with
$\eta^a$ and noting that $\eta^a{\widetilde{R}_a}{}^b=0$ in an
ultrastatic spacetime, that it must be $\lambda=0$.  Using again the
decomposition (\ref{dec}) and the property
$\widetilde{R}_{ab}\eta^b=0$, one finally finds that the Hobbs term
vanishes iff $v{\widetilde{R}_a}{}^b\tilde{\tau}_b=0$. Thus,
excluding the trivial case $v=0$, there is no local self-force for
uniform motion along optical geodesics whose directions are
eigenvectors of ${\widetilde{R}_a}{}^b$ with zero eigenvalue.  Such
directions exist iff the determinant of the matrix made by the
components ${\widetilde{R}_i}{}^j$ with $i,j=1,2,3$ is zero,
i.e., iff such matrix is degenerate.%

Let us reconsider the examples of Sec.\ \ref{sec2} in the light of
this conclusion.  For the Einstein universe,
${\widetilde{R}_i}{}^j={R_i}{}^j=(2/R^2){\delta_i}{}^j$, which is
always non-degenerate. For Schwarz\-schild spacetime,%
\begin{equation}%
\widetilde{R}_{\varphi
i}=\frac{2M}{r}\left(1-\frac{2M}{r}\right)^{-1}\left(1-
\frac{M}{r}\left(2+\sin^2\theta\right)\right)\delta_i^\varphi%
\end{equation}%
and, for $\theta=\pi/2$, ${\widetilde{R}_i}{}^j$ becomes degenerate
at $r=3M$.  Hence, $F_a^{({\rm l})}$ vanishes at $r=3M$ for uniform
motion in an equatorial plane along {\em any\/} optical geodesic,
not only for the circular orbit considered in Sec.\ \ref{sec2b}.
This result is far from trivial, but its derivation is rather
straightforward working in the optical space.%


\subsection{Geodesic motion}%
\label{sec3c}%

For charges following a geodesic in $({\cal M},\g_{ab})$ the
four-acceleration $a_a$ vanishes, so the von Laue term $F_a^{({\rm
vL})}$ is identically zero.\footnote{We avoid referring to this
situation as ``free fall,'' because the latter would be
appropriately identified by the vanishing of the external force in
Eq.\ (\ref{eqmot}), $K_a=0$, whereas geodesic motion corresponds to
$K_a+F_a=0$.} In optical geometry this result is not obvious,
because it follows from a compensation between contributions coming
from the ``optical'' von Laue and Hobbs terms. More precisely,
applying (\ref{relationship}) to this case one immediately finds
$\tilde{a}_a= -{k_a}^b\nab_b\Phi$.  Then, the von Laue term
$\widetilde{F}^{({\rm vL})}_a$ in optical space contains first and
second derivatives of $\Phi$, that are canceled by identical
contributions coming from the Hobbs term $\widetilde{F}^{({\rm
H})}_a$, because of the way the Ricci tensor
changes under a conformal transformation \cite{s98}.%

Thus, from the perspective of optical geometry, the fact that
$F_a^{({\rm vL})}=0$ just when $\tilde{a}_a=-{k_a}^b\nab_b\Phi$ is
somewhat surprising. For example, it is not evident that
$\widetilde{F}^{({\rm l})}_a=0$ for the Keplerian motions of Sec.\
\ref{sec2b}, because in optical geometry, where gravity is described
by the physical field $\Phi$, Keplerian motions have nothing
qualitatively different from, say, orbits in a Coulomb field. In
fact, it turns out that it is only thanks to a cancellation between
the von Laue and the Hobbs terms in the optical geometry that
$\widetilde{F}^{({\rm l})}_a=0$. However, if one thinks that for a
Keplerian motion in Schwarz\-schild spacetime $F^{({\rm
vL})}_a=F^{({\rm H})}_a=0$, and that $\widetilde{F}^{({\rm
l})}_a={\rm e}^\Phi F^{({\rm l})}_a$, then such a compensation is
only expected. Geodesics in $({\cal M},\g_{ab})$ are not geodesics
in $({\cal M},\tg_{ab})$, so $\widetilde{F}^{({\rm vL})}_a$ does not
vanish in general, and since $\widetilde{R}_{ab}\neq 0$, the Hobbs
force $\widetilde{F}^{({\rm H})}_a$ also does not vanish.
Nevertheless, $\widetilde{F}^{({\rm vL})}_a$ and
$\widetilde{F}^{({\rm H})}_a$ combine to form a vanishing
local self-force, $\widetilde{F}^{({\rm l})}_a=0$.%

This remark should not be regarded as a drawback of the description
based on optical geometry, though.  Indeed, the fact that
$F_a^{({\rm vL})}$ does, or does not, vanish is not particularly
important, because the physically interesting quantity is not
$F_a^{({\rm vL})}$, but $F_a^{({\rm l})}$. And whether $F_a^{({\rm
l})}=0$ (or, equivalently, $\widetilde{F}_a^{({\rm l})}=0$) cannot
be established simply by looking at $F_a^{({\rm vL})}$ (or
$\widetilde{F}^{({\rm vL})}_a$), unless one is in a situation for
which $F^{({\rm H})}_a=0$ (or $\widetilde{F}^{({\rm H})}_a=0$).
Thus, in the case of Schwarz\-schild spacetime the evaluation of
$F_a^{({\rm vL})}$ turns out to be convenient only because, since
$R_{ab}=0$ there, the Hobbs force vanishes identically, so
$F_a^{({\rm vL})}$ coincides with $F_a^{({\rm l})}$.  However, in a
spacetime for which it is $\widetilde{R}_{ab}$ that vanishes, one
has $\widetilde{F}^{({\rm H})}_a=0$.  Consequently, it is now
$\widetilde{F}^{({\rm vL})}_a$ that one should inspect, in order to
get information about the behaviour of $F_a^{({\rm l})}=0$.%


\subsection{Conformally static charge}%
\label{sec3d}%

In order to further illustrate the point made in the last paragraph
of Sec.\ \ref{sec3c}, let us consider a conformally static charge,
i.e., one with four-velocity $v^a=n^a$, where%
\begin{equation}%
n^a:=\ee^{-\Phi}\eta^a
=\left(-\g_{bc}\eta^b\eta^c\right)^{-1/2}\eta^a%
\label{n}%
\end{equation}%
is the unit vector field parallel to $\eta^a$.  This is obviously a
subcase of the situations covered in Sec.\ \ref{sec3a}, and since
$\widetilde{R}_{ab}\eta^b=0$, it is evident in optical geometry that
$\widetilde{F}^{({\rm vL})}_a =\widetilde{F}^{({\rm H})}_a=0$, so
also $\widetilde{F}^{({\rm l})}_a=0$ and $F^{({\rm l})}_a=0$.  On
the other hand, this conclusion is not obvious at all if one works
in the spacetime $({\cal M},\g_{ab})$, because in order to establish
it, one must use highly non-trivial properties of the vector field
$n^a$, that lead to cancelations between terms coming from
$F_a^{({\rm vL})}$ and $F_a^{({\rm H})}$ (none of which, however,
vanishes separately).  Hence, in this case it is optical geometry
that allows one to establish that $F_a^{({\rm l})}=0$ in a simple
way, by using the decomposition $\widetilde{F}_a^{({\rm
l})}=\widetilde{F}_a^{({\rm vL})}+\widetilde{F}_a^{({\rm H})}$.  The
alternative split, $F_a^{({\rm l})}=F_a^{({\rm vL})}+F_a^{({\rm
H})}$, leads instead to rather cumbersome calculations.%


\section{Conclusions}%
\label{sec4}%
\setcounter{equation}{0}

Let us summarise the results of Secs.\ \ref{sec2}--\ref{sec3}.  We
have seen that, for uniform motion along a spatial geodesic
$\chi=\pi/2$ in Einstein's static universe, the von Laue force
vanishes, whereas the Hobbs force does not (hence, the local part of
the self-force also does not vanish).  In Schwarz\-schild spacetime,
for uniform motion along a circle at $r=3M$, the von Laue force
vanishes; since the Hobbs force is identically zero (Schwarz\-schild
spacetime is Ricci-flat), this implies that $F^{({\rm l})}_a(3M)=0$.
Similarly, in the optical Schwarz\-schild spacetime,
$\widetilde{F}^{({\rm vL})}_a(3M)=\widetilde{F}^{({\rm
H})}_a(3M)=\widetilde{F}^{({\rm l})}_a(3M)=0$.  All these are
instances of the following general results, valid in an arbitrary
conformally static spacetime: (i) for geodesic motion in optical
spacetime $({\cal M},\tg_{ab})$ --- uniform motion along optical
geodesics ---, $\widetilde{F}^{({\rm vL})}_a=0$; (ii) for geodesic
motion in the spacetime $({\cal M},\g_{ab})$, $F^{({\rm vL})}_a=0$.
Note that, however, the physically interesting result does not
concern $F^{({\rm vL})}_a$ or $\widetilde{F}^{({\rm vL})}_a$, but
the entire local part of the self-force.  Whether this vanishes in
cases (i) and (ii) depends on the additional feature that the matrix
${\widetilde{R}_i}{}^j$ or $R_{\mu}{}^\nu$, respectively, be
degenerate.  Also, note that the particular case of (i), that for a
conformally static charge $F^{({\rm l})}_a=\widetilde{F}^{({\rm
l})}_a=0$, although non-trivial, is straightforward when regarded
from the perspective offered by optical geometry.  Hence, the latter
appears to be a useful tool in calculations of self-force and
related effects, as it happens already in several other
circumstances
\cite{acl,a93,a90,rev,torres,saa,perlick,ams,peri,maxwell}. Indeed,
although in this paper we deliberately avoided dealing with the
non-local part of the self-force $F^{({\rm nl})}_a$, there are good
reasons to believe that working in the optical spacetime might also
simplify its evaluation --- a rather challenging task, in general.
This is suggested by the fact that computing $F^{({\rm nl})}_a$
amounts, basically, to finding the bi-tensor $f_{aba'}$ in Eq.\
(\ref{nonloc}).  In turn, this amounts to the determination of a
particular electromagnetic field in spacetime, and we know already
from Ref.\ \cite{maxwell} that this is sometimes easier to do using
optical geometry.  We leave the development of this
subject for future investigations.%


\section*{Acknowledgements}%
It is a pleasure to thank Rossella Rosin for her careful preparation
of the figures.  S.S.\ is grateful to Nordita and to the Department
of Astronomy and Astrophysics at Chalmers University for
hospitality.%


\setcounter{section}{0}%
\def\thesection{\Alph{section}}%

\section*{Appendix: Centripetal acceleration in
optical geometry}%

\def\theequation{A\arabic{equation}}%
\label{appA}%
\setcounter{equation}{0}%

We present a derivation of Eq.\ (\ref{optacc}) based on the
embedding diagram of Fig.\ \ref{fig1}. In the fictitious
three-dimensional Euclidean space, the centripetal acceleration of a
particle that moves uniformly on the orbit $r=\mbox{const}$ is
$a_{\scriptscriptstyle{\rm E}}^{{\rm (c)}}=\gamma^2\Omega^2\rho$,
where $\rho=\left(1-2M/r\right)^{-1/2}r$ is the distance from the
axis \cite{ksa} --- also called ``radius of gyration'' \cite{ams}
because $v=\left(1-2M/r\right)^{-1/2}\Omega r=\Omega\rho$. The
centripetal acceleration in the optical space is given by the
tangential component of $a_{\scriptscriptstyle{\rm E}}$, namely%
\begin{equation}%
a^{{\rm (c)}}=\gamma^2\Omega^2\rho\cos\alpha\;,%
\label{cos}%
\end{equation}%
where $\alpha$ is the angle indicated in Fig.\ \ref{fig1}. Let
$z=f(\rho)$ be the equation of the surface in Fig.\ \ref{fig1}. The
surface is isometric to the section $\theta=\pi/2$ of the
optical space if \cite{ksa}%
\begin{equation}%
\left(1+\left(\frac{\dd
f}{\dd\rho}\right)^2\right)\dd\rho^2=
\left(1-\frac{2M}{r}\right)^{-2}\dd r^2\;.%
\end{equation}%
But we have also $\dd f/\dd\rho=\tan\alpha$, so%
\begin{equation}%
\cos\alpha=\left(1-\frac{2M}{r}\right)
\left|\frac{\dd\rho}{\dd
r}\right|=\left(1-\frac{2M}{r}\right)^{-1/2}\left|1-
\frac{3M}{r}\right|\;.%
\end{equation}%
Substituting into (\ref{cos}) we recover Eq.\ (\ref{optacc}).%


%

\end{document}